\def\bey{\begin{eqnarray}}
\def\eey{\end{eqnarray}}
\def\be{\begin{equation}}
\def\ee{\end{equation}}
\def\ba{\begin{array}}
\def\ea{\end{array}}

\documentclass[onecolumn,showpacs,preprintnumbers,amsmath,amssymb]{revtex4}
\usepackage{graphicx}
\usepackage{fancyhdr}
\usepackage{dcolumn}
\usepackage{bm}
\begin{document}
\preprint{ }
\title
{\Large Warm strange hadronic matter in an effective model with a
weak Y-Y interaction}
\author{\large  W.L.Qian$^1$\footnote{wlqian@fudan.edu.cn}, R.K. Su$^{2,1,4}$ \footnote{rksu@fudan.ac.cn}
and H.Q.Song$^{2,3,4}$\footnote{songhq@sinr.ac.cn}}

\vspace{10mm}

 \affiliation{
\small
1. Department of Physics, Fudan University, Shanghai 200433,
China\\
 \small 2.
CCAST(World Laboratory), P.O.Box 8730, Beijing 100080, China\\
\small
3. Shanghai Institute of Applied Physics, Chinese Academy of
Sciences, P.O.Box 800204, Shanghai 201800, China\\
 \small 4.
Research Center of Nuclear Theory of National Laboratory of Heavy
Ion \\ \small Accelerator of Lanzhou, Lanzhou 730000.}

\baselineskip 20.6pt
\begin{abstract}
An effective model is used to study the equation of state(EOS) of
warm strange hadronic matter with nucleons, $\Lambda$-hyperons, $\Xi$-hyperons,
$\sigma^{*}$ and $\phi$. In the calculation, a newest weak Y-Y interaction
deduced from the recent observation\cite{Takahashi} of a
$_{\Lambda\Lambda}^6$He double hypernucleus is adopted.
Employing this effective model, the results with strong Y-Y interaction and weak
Y-Y interaction are compared.
\end{abstract}


\maketitle
\newpage
\newpage
\section{\bf\large {Introduction}}

Normal nuclei are made of protons and neutrons. If hyperon carrying strangeness
is added into a nucleus somehow, we
then obtain hypernuclei.  The first hypernucleus
was seen in emulsion by Danysz and
Pniewski\cite{Danysz1} in 1953.  Since then, strangeness carried
by s-quark opens a new dimension for the studies in nuclear
physics. In recent years, exploring nuclear system with
strangeness, i.e., strange matter has received increasing
interest. Such system has many astrophysical and cosmological
implications and is indeed interesting by itself. For instance,
the core of neutron stars may contain a high fraction of
hyperons\cite{Schaffner,Yamamoto,Sahu}, resulting in a third
family of compact stars which has a similar mass to neutron star
but has a much smaller radius than the later\cite{Schaffner}. The
strange matter may also be formed in relativistic heavy ion
collisions. There are two kinds of strange matter: strange quark
matter and strange hadronic matter. On one hand, it has been
speculated\cite{Witten,Bodmer,Chin,Jaffe} that lumps of quark
matter ("strangelets") with large strangeness per baryon might be
more stable than the normal nuclei. The experimental work
searching for strange quark matter has been going on  in BNL-AGS
and CERN-SPS \cite{Armstrong,Appelquist,Arsenescu,Hill}.
But unfortunately, the evidence for the production of strangelets has
not yet been
observed within the experimental limits. On the other hand,
strange hadronic matter or hypernuclei have also been investigated
\cite{Ikeda,Barranco,Schaffner1,Schaffner2,Schaffner3,Schulze,Vidana,Zhang1,Wang1,Zhang2,Wang2,Wang3}.
But the inclusion of multiple units of strangeness in nuclei remains
rather largely unexplored yet. This is because the technical
difficulty experimentally and the uncertainty of the interactions
between baryons theoretically.  Recently, Takahashi et
al.\cite{Takahashi} reported their observation of a
$_{\Lambda\Lambda}^6$He double hypernucleus, where the
$\Lambda-\Lambda$ interaction energy $\Delta
B_{\Lambda\Lambda}=1.01\pm0.20^{+0.18}_{-0.11}$MeV is deduced from
measured data.  This value is much smaller than the previous
estimation $\Delta B_{\Lambda\Lambda}\simeq 4-5$MeV from the early
experiments\cite{Danysz2,Prowse,Aoki,Song}.
To incorporate the early strong Y-Y interaction, the $\Lambda$ well depth in
"$\Lambda$ matter" at density $0.5\rho_0$ was estimated as
$V_\Lambda^{(\Lambda)}\simeq 20$ MeV by
using the Nijmegen model D.\cite{Schaffner3}. If the new value $\Delta
B_{\Lambda\Lambda}=1.01$MeV is used, the
$V_\Lambda^{(\Lambda)}\simeq 5$ MeV is obtained.
The measurement
of $_{\Lambda\Lambda}^4H$ was also performed in BNL, but
they did not obtain the interaction between the lambdas.

Although QCD is the fundamental theory for strong interaction,
it is not available to describe strange hadronic matter directly
because of its non-perturbative properties. Two kinds of effective
models had been introduced.
The first kind is focused on the chiral SU(3) symmetry\cite{Papazoglou1,Papazoglou2}.
For example, in ref. \cite{Papazoglou1}, a generalized Lagrangian which is
based on the linear realization of chiral SU(3) symmetry and the concept of
broken scale invariance was proposed to describe the SHM.
The second kind is based upon the successful models of nuclear matter,
for example, Walecka model\cite{Schaffner2,Schaffner3,Schaffner4},
quark-meson coupling(QMC) model\cite{Saito}, or Furnstahl-Serot-Tang(FST) model\cite{FST},
and adding hyperons into these models.
In refs.\cite{Schaffner2} and \cite{Schaffner3}, two models
(denoted as model 1 and model 2)
which are based on a SU(3) extension of the Walecka model are suggested
to deal with the weak and strong Y-Y interaction.
In particular, in the model 2, in order to incorporate the early strong Y-Y interaction data\cite{Danysz2},
two additional mesons, namely, $\sigma^*$ and $\phi$ mesons are introduced to
obtain the strong attraction between $\Lambda$ hyperons.
The advantage of this treatment is that
it can easily reduce to an effective model which can explain the nuclear systems
very well when the hyperons are taken away, because the coupling constants between
nucleons and mesons are determined by the experimental data of nuclear system.
In this sense, the effect of hyperons can be exposed explicitly.
Of course, part of the second kind of models suffer from the loss of chiral SU(3) symmetry.
In fact, the treatment for adding hyperons to nuclear system to study the hypernuclei,
for example, $\Lambda$ hypernuclei has been employed by many previous papers.

Recently, by using an extended modified quark meson coupling(MQMC)
model we studied the implications\cite{Song} of the newest Y-Y
interaction in strange hadronic matter. It is found that  while
the system with the strong $\Lambda-\Lambda$ interaction and in a
quite large strangeness fraction region is more deeply bound than
the ordinary nuclear matter due to the opening of the new degrees
of freedom, the system with the weak $\Lambda-\Lambda$ interaction
is rather loosely bound compared to the later.  It is interesting
to check if the above remarks depend on the model used and to see
what happens if the system is at finite temperature. In a previous
paper\cite{Zhang2} , we suggested an effective model, constructed
by introducing hyperons in the Furnstahl-Serot-Tang(FST)
model, to study the saturation properties and
stabilities of strange hadronic matter at both zero\cite{Zhang2}
and finite temperature\cite{Qian}. In this work, we will use the
extended FST model to study warm strange hadronic matter with the
newest weak Y-Y interactions and then compare the results to
those\cite{Qian} with previous strong Y-Y interactions. The
details of this model can be found in ref.\cite{Zhang2}, here we
only give a short
description. Considering reactions, $\Lambda +\Lambda \rightarrow \Xi ^{-}+p$,
$\Lambda +\Lambda \rightarrow \Xi ^0+n$ and their reverses, we
take the mixture of the cascades $\Xi ^{-}$ and $\Xi ^0$ in the
strange matter into our model, besides lambdas. For simplicity, we
assume that $\Xi ^{-}$ and $\Xi ^0$ will appear in the strange
matter with equal amount. This is similar to the protons and
neutrons in symmetric nuclear matter. We used, therefore, a single
symbol $\Xi $ for these particles. Furthermore, we have also
include the $\sigma ^{*}$ and $\phi $ mesons in the model to
describe the interaction between hyperons, as proposed by
Schaffner et al.\cite {Schaffner3}.
 We will not consider the
mixture of the $\Sigma$ hyperons in the same reason as mentioned
in Ref.\cite{Zhang2}. The paper is organized as follows. In
section II, we will gives a brief description of the model. The
calculated results and some discussions will be presented in
section III.

\section{The extended FST model}
In our previous paper \cite{Zhang2,Qian}, the original FST
model\cite{FST} was extended by including $\Lambda$
 and $\Xi$ hyperons in the system and introducing a new
hyperon-hyperon interaction mediated by two additional strange
mesons $\sigma^*$ and $\phi$ which couple  only to hyperons, just
as we stated in the introduction section. Since we will study the
unpolarized system, the $\pi$ meson has no influence on the
system. Omitting the contributions from the $\pi$ meson, we have
  the following Lagrangian density
for the FST model.
\begin{eqnarray}
{\cal L}(x) &=&\bar{\psi}_N(i\gamma ^\mu \partial _\mu -g_{\omega
N}\gamma ^\mu V_\mu -M_N+g_{sN}\sigma )\psi _N  \nonumber \\
&&+\bar{\psi}_\Lambda (i\gamma ^\mu {\partial }_\mu -g_{\omega
\Lambda }\gamma ^\mu V_\mu -g_{\phi \Lambda }\gamma ^\mu \phi _\mu
-M_\Lambda +g_{s\Lambda }\sigma +g_{\sigma ^{*}\Lambda }\sigma
^{*})\psi _\Lambda \nonumber \\ &&+\bar{\psi}_\Xi (i\gamma ^\mu
{\partial }_\mu -g_{\omega \Xi }\gamma ^\mu V_\mu -g_{\phi \Xi
}\gamma ^\mu \phi _\mu -M_\Xi +g_{s\Xi }\sigma +g_{\sigma ^{*}\Xi
}\sigma ^{*})\psi _\Xi  \nonumber \\ &&-\frac 14G_{\mu \nu }G^{\nu
\mu }+\frac 12\left( 1+\eta \frac \sigma {S_0}\right) m_\omega
^2V_\mu V^\mu +\frac 1{4!}\zeta \left( g_\omega ^2V_\mu V^\mu
\right) ^2  \nonumber \\ &&+\frac 12\partial _\mu \sigma \partial
^\mu \sigma -H_q\left( \frac{S^2}{ S_0^2}\right) ^{2/d}\left(
\frac 1{2d}ln\frac{S^2}{S_0^2}-\frac 14\right) \nonumber \\
&&-\frac 14S_{\mu \nu }S^{\mu \nu }+\frac 12m_\phi ^2\phi _\mu
\phi ^\mu +\frac 12\left( \partial _\nu \sigma ^{*}\partial ^\nu
\sigma ^{*}-m_{\sigma ^{*}}^2\sigma ^{*^2}\right) .
\end{eqnarray}
where $g_{ij}$ are the coupling constants of the baryons to the
meson fields. $G_{\mu \nu }=\partial _\mu V_\nu -\partial _\nu
V_\mu $ and $S_{\mu \nu }=\partial _\mu \phi _\nu -\partial _\nu
\phi _\mu $ are the $\omega $ field and the $\phi $ field strength
tensor, respectively. The scalar fluctuation field $\sigma $ is
related to $S$ by $S(x)\equiv S_0-\sigma (x)$
. $H_q$ is linked to the mass of the light scalar $S$ by the relation $%
m_s^2=4H_q/(d^2S_0^2)$. For the symmetric matter considered here,
there is no contribution from $\rho $ meson field. In mean field
approximation, the Lagrangian density then takes the form
\begin{eqnarray}
{\cal L}_{MFT} &=&\bar{\psi}_N(i\gamma ^\mu \partial _\mu
+g_{\omega N}\gamma ^0V_0-M_N+g_{sN}\sigma _0)\psi _N  \nonumber
\\ &&+\bar{\psi}_\Lambda (i\gamma ^\mu {\partial }_\mu -g_{\omega
\Lambda }\gamma ^0V_0-g_{\phi \Lambda }\gamma ^0\phi _0-M_\Lambda
+g_{s\Lambda }\sigma _0+g_{\sigma ^{*}\Lambda }\sigma _0^{*})\psi
_\Lambda  \nonumber \\ &&+\bar{\psi}_\Xi (i\gamma ^\mu {\partial
}_\mu -g_{\omega \Xi }\gamma ^0V_0-g_{\phi \Xi }\gamma ^0\phi
_0-M_\Xi +g_{s\Xi }\sigma _0+g_{\sigma ^{*}_0\Xi }\sigma
_0^{*})\psi _\Xi  \nonumber \\ &&+\frac 12\left( 1+\eta
\frac{\sigma _0}{S_0}\right) m_\omega ^2V_0^2+\frac 1{4!}\zeta
\left( g_{\omega N}V_0\right) ^4+\frac 12m_\phi ^2\phi _0^2-\frac
12m_{\sigma ^{*}}^2\sigma_0^{*^2}.  \nonumber \\
&&-H_q\left( 1-\frac{\sigma _0}{S_0}\right) ^{4/d}\left[ \frac 1dln\left( 1-%
\frac{\sigma _0}{S_0}\right) -\frac 14\right] ,
\end{eqnarray}
where the meson field operators are replaced by their mean field values: $%
\phi _0$, $V_0$ , $\sigma _0$ and $\sigma _0^{*}$. To derive the
equation of
state at finite temperature, we calculate the thermodynamic potential $%
\Omega $ by using the standard technique in the field theory and
statistical mechanics. The result reads
\begin{eqnarray}
\Omega &=&V\{H_g[(1-\frac{\phi _0}{S_0})^{\frac 4d}(\frac 1d\ln
(1-\frac{ \phi _0}{S_0})-\frac 14)+\frac 14]  \nonumber \\
&&-\frac 12(1+\eta \frac{\phi _o}{S_0})m_\omega^2V_0^2-\frac
1{4!}\zeta (g_{\omega N}V_0)^4-\frac 12m_\phi ^2\phi _0^2+\frac
12m_{\sigma ^{*}}\sigma_0 ^{*2}\}  \nonumber \\
&&-2k_BT\{\sum_{i,{\bf k}}\ln {[1+e^{-\beta (E_i^{*}(k)-\nu
_i}]}+\sum_{{i, {\bf k}}}\ln {[1+e^{-\beta (E_i^{*}(k)+\nu
_i)}]}\},
\end{eqnarray}
where $\beta =1/k_BT$ and $V$ is the volume of the system.
\begin{equation}
E_i^{*}(k)=\sqrt{M_i^{*2}+k^2}
\end{equation}
with
\begin{equation}
M_i^{*}=M_i-g_{si}\sigma _0-g_{\sigma ^{*}i}\sigma
_0^{*}(i=\Lambda ,\Xi ),
\end{equation}
\begin{equation}
M_i^{*}=M_i-g_{si}\sigma _0(i=N)
\end{equation}
being the effective baryon masses.The chemical potential $\nu _i $
is calculated from baryon density $\rho_{Bi}$ by the subsidiary
conditions
\begin{equation}
\rho _{Bi}=\frac 2{(2\pi )^3}\int
d^3k[n_i(k)-\overline{n}_i(k)],\hspace{2cm} (i=n,p,\Lambda ,\Xi
^0,\Xi ^{-})
\end{equation}
where $n_i(k)$ and $\overline{n}_i(k)$ are baryon and anti-baryon
distributions and expressed as
\begin{equation}
n_i(k)=\{exp[(E_i^{*}(k)-\nu _i)/k_BT]+1\}^{-1}
\end{equation}
and
\begin{equation}
\overline{n}_i(k)=\{exp[(E_i^{*}(k)+\nu
_i)/k_BT]+1\}^{-1}.\hspace{2cm}
\end{equation}

Having obtained the thermodynamic potential, one can easily
calculate all other thermodynamic quantities of the system. For
example, the pressure $p$ is given by the relation $p=-\Omega /V$,
the average energy density  ${\cal E}$
 by ${\cal E} V=\partial
(\beta \Omega )/\partial \beta +\mu \rho V $. The resulting
expressions are as the follows.
\begin{eqnarray}
{\cal E} &=&\frac 2{(2\pi )^3}\sum_i\int
d^3kE_i^{*}(k)[n_i(k)+\overline{n} _i(k)]+H_q\left\{ \left(
1-\frac{\phi _0}{S_0}\right) ^{\frac 4d}\left[ \frac 1dln\left(
1-\frac{\phi _0}{S_0}\right) -\frac 14\right] +\frac 14\right\}
\nonumber \\ &&+g_{\omega N}V_0\rho _{BN}+\left( g_{\omega \Lambda
}V_0+g_{\phi \Lambda }\phi _0\right) \rho _{B\Lambda }+\left(
g_{\omega \Xi }V_0+g_{\phi \Xi }\phi _0\right) \rho _{B\Xi }
\nonumber \\ &&-\frac 12\left( 1+\eta \frac{\phi _0}{S_0}\right)
m_\omega^2V_0^2-\frac 1{4!}\zeta g_{\omega N}^4V_0^4-\frac
12m_\phi ^2\phi _0^2+\frac 12m_{\sigma ^{*}}^2\sigma _0^{*^2}
\end{eqnarray}
and
\begin{eqnarray}
p &=&\frac 13\frac 2{(2\pi )^3}\sum_i\int
d^3k\frac{k^2}{E_i^{*}(k)}[n_i(k)+ \overline{n}_i(k)]-H_q\left\{
\left( 1-\frac{\phi _0}{S_0}\right) ^{\frac 4d}\left[ \frac
1dln\left( 1-\frac{\phi _0}{S_0}\right) -\frac 14\right] +\frac
14\right\}  \nonumber \\ &&+\frac 12\left( 1+\eta \frac{\phi
_0}{S_0}\right) m_\omega^2V_0^2+\frac 1{4!}\zeta g_{\omega N}
^4V_0^4+\frac 12m_\phi ^2\phi _0^2-\frac 12m_{\sigma ^{*}}^2\sigma
_0^{*^2}.
\end{eqnarray}

\begin{table}[tbp]
\caption{Parameter sets, where values for $S_{0}$, the scalar mass
$m_{s}$ are in MeV.}
\begin{tabular}{ccccccccccccc}
\hline $g_{sN}^{2}$ & $m_{s}$ & $g_{\omega N}^{2}$ & $S_{0}$ &
$\zeta$ & $\eta$ & d & $g_{s\Lambda}^2$ & $g_{s\Xi}^2$ &
$g_{\sigma^*\Lambda}^2$(S) & $g_{\sigma^*\Xi}^2$(S) &
$g_{\sigma^*\Lambda}^2$(W)&$g_{\sigma^*\Xi}^2$(W)\\ \hline 99.3 &
509 & 154.5 & 90.6 & 0.0402 & -0.496 & 2.70 & 37.32 & 9.99 & 48.31
& 154.62 & 28.73 & 129.06\\ \hline
\end{tabular}
\end{table}

Similarly, Helmholtz free energy $F$ is calculated from internal
energy $E$ and entropy $S$ by
formula $F=E-TS$. Entropy $S$ is obtained from thermodynamic potential $%
\Omega$ by using relation $S=-(\partial \Omega /\partial
T)_{V,\mu}$.

Now we come to discuss the chemical equilibrium condition for the reactions $%
\Lambda +\Lambda \rightleftharpoons n+\Xi ^0$ and $\Lambda
+\Lambda \rightleftharpoons p+\Xi ^{-}$. As pointed out in Sec.I,
we will discuss the
system with equal number of protons and neutrons as well as equal number of $
\Xi ^0$ and $\Xi ^{-}$. In this case, the chemical equilibrium
condition reads
\begin{equation}
2\nu _\Lambda -\nu _N-\nu _\Xi =0.
\end{equation}
We defines a strangeness fraction $f_S$ as
\begin{equation}
f_S\equiv \frac{\rho _{B\Lambda }+2\rho _{B\Xi }}{\rho _B},
\end{equation}
where
\begin{equation}
 \rho _B=\rho _{BN}+\rho _{B\Lambda }+\rho _{B\Xi }.
 \end{equation}
  Given $\rho _B$
and $f_S$, we determine $\rho _{BN}$, $\rho _{B\Lambda }$ and
$\rho _{B\Xi }$ by above three equations. \bigskip

\section{Results and discussions}

In numerical calculation, we adopt the same parameters as in
Ref.\cite{Zhang2}(quoted in Table 1). The symbol S(W) in the
parentheses denotes these coupling constants are deduced from the
strong(weak) Y-Y interaction.  To determine the
coupling constant $g_{\sigma^*\Lambda}$ and
$g_{\sigma^*\Xi}$, we use the estimation made by Schaffner et. al.\cite{Schaffner3}.
By using the Nijmegen model D and the method given by Millener\cite{Millener},
they found
\begin{equation}
U_\Lambda^{(\Xi)} \simeq U_\Xi^{(\Xi)} \simeq 2U_\Lambda^{(\Lambda)}
\end{equation}
at densityes of $\rho_{\Xi}=\rho_0$ and $\rho_{\Lambda}=\rho_{0}/2$,
where the notation $U_{Y}^{(Y')}$ stands for the potential depth for
hyperon $Y$ in a "bath" of hyperon $Y'$, and
\begin{equation}
\frac {U_\Lambda^{(\Lambda)}} {U_{N}^{(N)}} = \frac{1}{2} \frac{1/4 V_{\Lambda \Lambda}}{3/8 V_{NN}}
\end{equation}
For strong Y-Y interactions, $V_{\Lambda \Lambda} \equiv \Delta B _{\Lambda\Lambda}\simeq 4-5 MeV$
and $V_{NN} \simeq 6-7 MeV$, we have $V_{\Lambda \Lambda}/V_{NN} \simeq 3/4$.
In relativistic mean field, $U_{N}^{(N)} \simeq 80 MeV$, we obtain
$U_\Lambda^{(\Xi)} \simeq U_\Xi^{(\Xi)} \simeq 40 MeV$.
But for weak Y-Y interaction, $\Delta B _{\Lambda\Lambda}\simeq 1.01 MeV$,
we have
$U_\Lambda^{(\Xi)} \simeq U_\Xi^{(\Xi)} \simeq 10 MeV$,
and $g_{\sigma^*\Lambda}(W) = 5.46$, $g_{\sigma^*\Xi}(W) = 11.39$.

Besides these, we set
$g_{\omega\Lambda}/g_{ \omega N}=2/3$, $g_{\omega\Xi}/g_{\omega
N}=1/3$, according to the OZI rule \cite{Dover1} and used the
quark model relationships $g_{\phi \Xi} = 2g_{\phi \Lambda}=
-2\sqrt{2}g_{\omega N}/3$. The bare masses of baryons and
mesons are $M_{N}=939$ MeV ,$M_{\Lambda}=1116$ MeV ,$M_{\Xi}=1318.1$ MeV,
$m_{\omega}=783$ MeV, $m_{\sigma^{*}}=975$ MeV and $m_{\phi}=1020$
MeV.

\begin{figure}
\begin{center}
\includegraphics[width=7cm,height=10cm,angle=0]{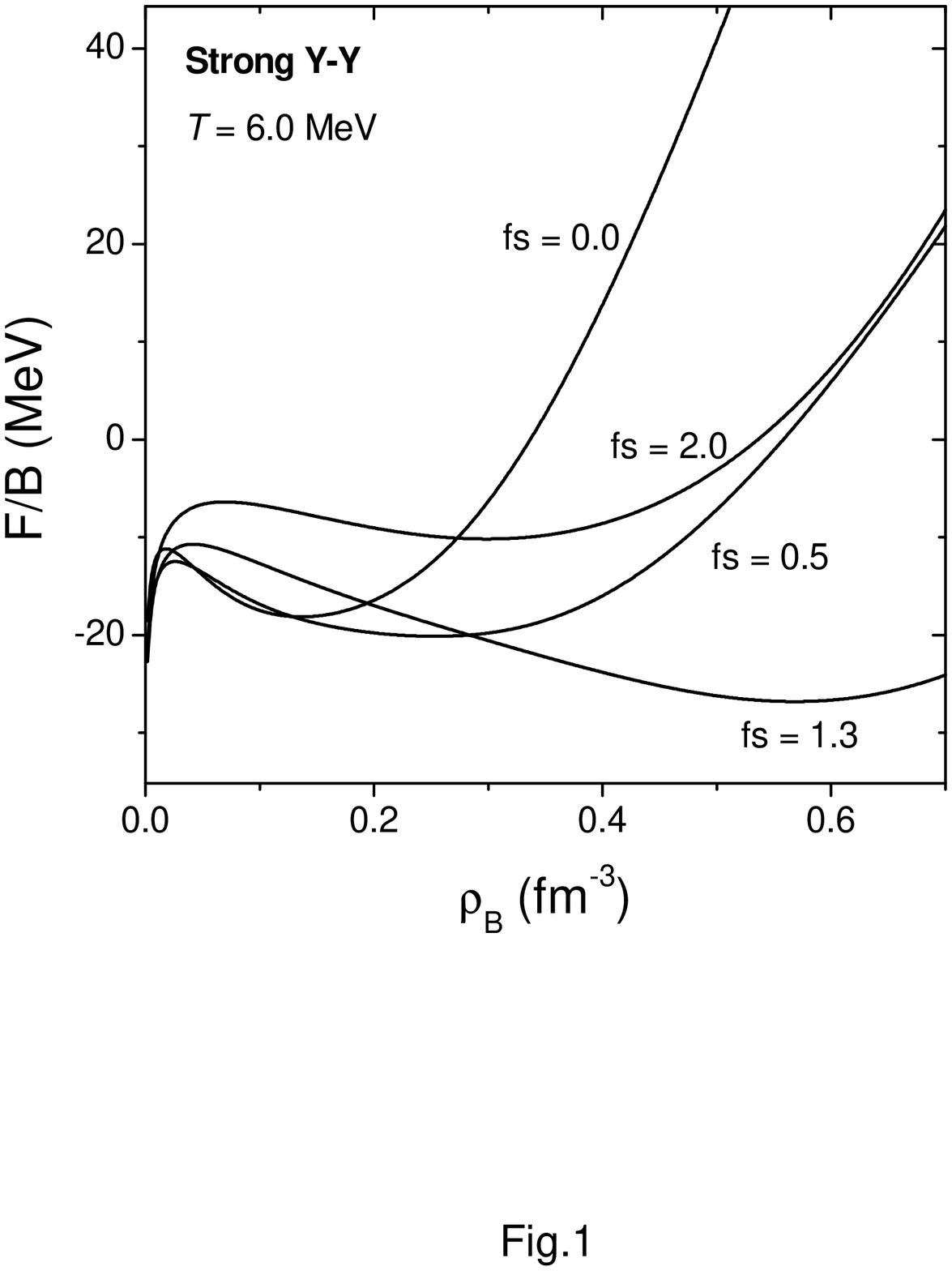}
 \end{center}
\caption{Free energy per baryon versus baryon density in the
strange hadronic matter with various values of strangeness
fraction and at temperature $T=6MeV$, calculated with a strong
$\Lambda-\Lambda$ interaction.\label{fig1} }
\end{figure}

We will discuss the Helmholtz free energy $F$ of the strange
hadronic matter first.   As usually, we subtract the baryon masses
in the free energy per baryon of the strange matter.
\begin{equation}
{\frac{F }{B}}=({\frac{F
}{B}})_{tot}-M_N(1-Y_\Lambda-Y_\Xi)-M_\Lambda Y_\Lambda-M_\Xi
Y_\Xi.
\end{equation}
where $Y_{\Lambda}=\rho_{B\Lambda}/\rho_{B}$ and
$Y_{\Xi}=\rho_{B\Xi}/ \rho_{B}$ are the hyperon fractions in the
matter. In Fig.1, we have plotted the free energy per baryon
versus the baryon density of the matter with various strangeness
fractions at temperature $T=6MeV$, calculated with the old strong
Y-Y interactions. The outstanding feature is that with the
increasing strangeness fraction, the saturation curves become
deeper first and then shallower. The lowest minimum occurs around
$f_{S}=1.3$. The corresponding saturation density increases from
0.148${fm}^{-3}$ for ordinary nuclear matter($fs=0$) to a maximum
value $\sim 0.56{fm}^{-3}$ at $f_S \simeq 1.3$ and then decreases.
We also note that the F/B curve for any value of $f_S$ has a
negative minimum. It means that systems at $T=6MeV$ and with any
combination satisfying constraint Eq.(12) will be stable against
particle emission. The Fig.2 shows the same curves as in Fig.1 but
calculated with the newest weak Y-Y interactions.  Due to the
weakness of the Y-Y interactions used in this case, the saturation
curves become shallower and shallower with increasing strangeness
fraction $fs$ except at very small $fs$ value around 0.1. For the
$fs$ value larger than about 1.25, there is no negative minimum in
the curve.  It means that the system will no longer be stable in
this region.

\begin{figure}
\begin{center}
\includegraphics[width=7cm,height=10cm,angle=0]{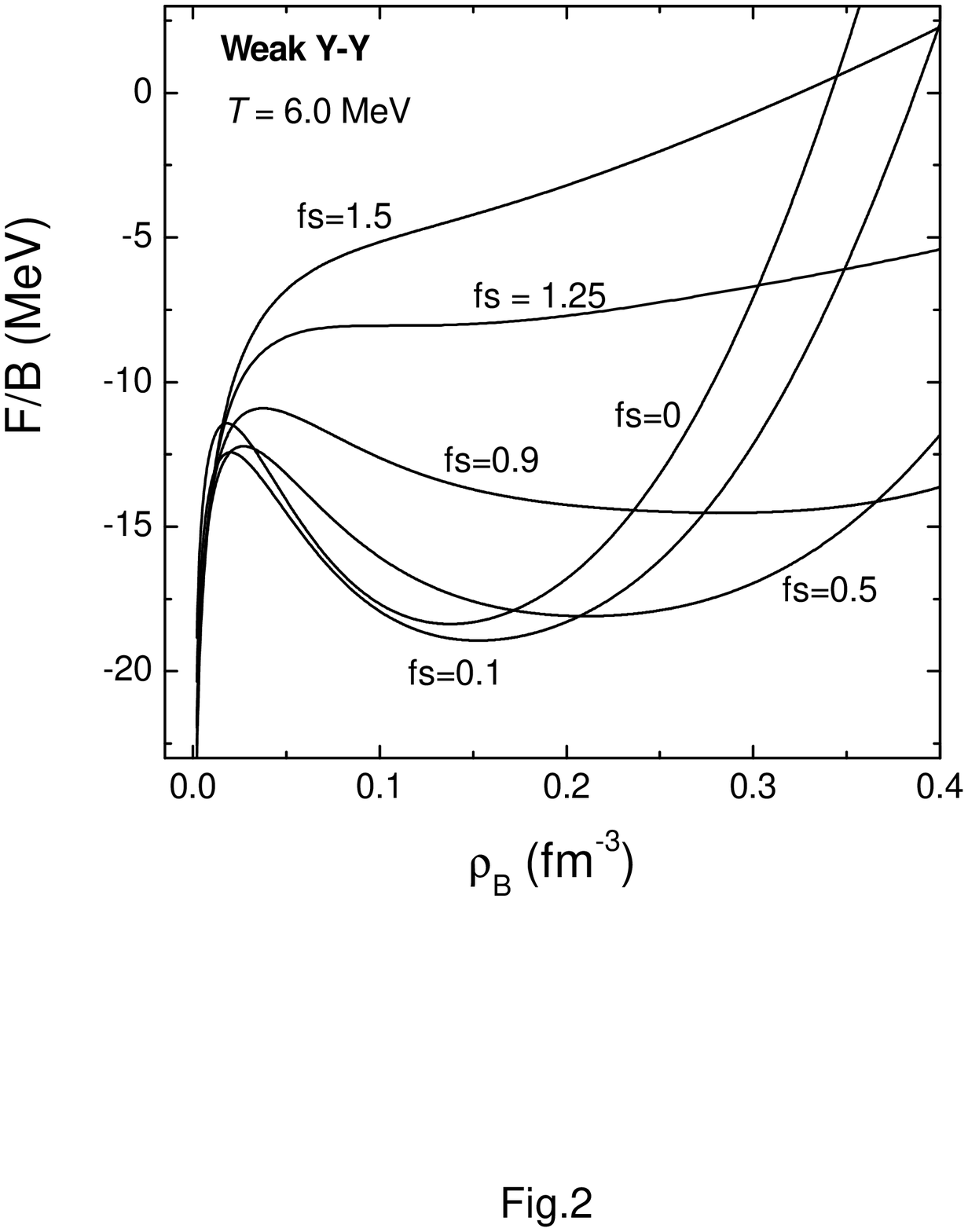}
 \end{center}
\caption{The same curves as in Fig.1 but calculated with a weak
$\Lambda-\Lambda$ interaction.\label{fig2} }
\end{figure}

\begin{figure}
\begin{center}
\includegraphics[width=7cm,height=10cm,angle=0]{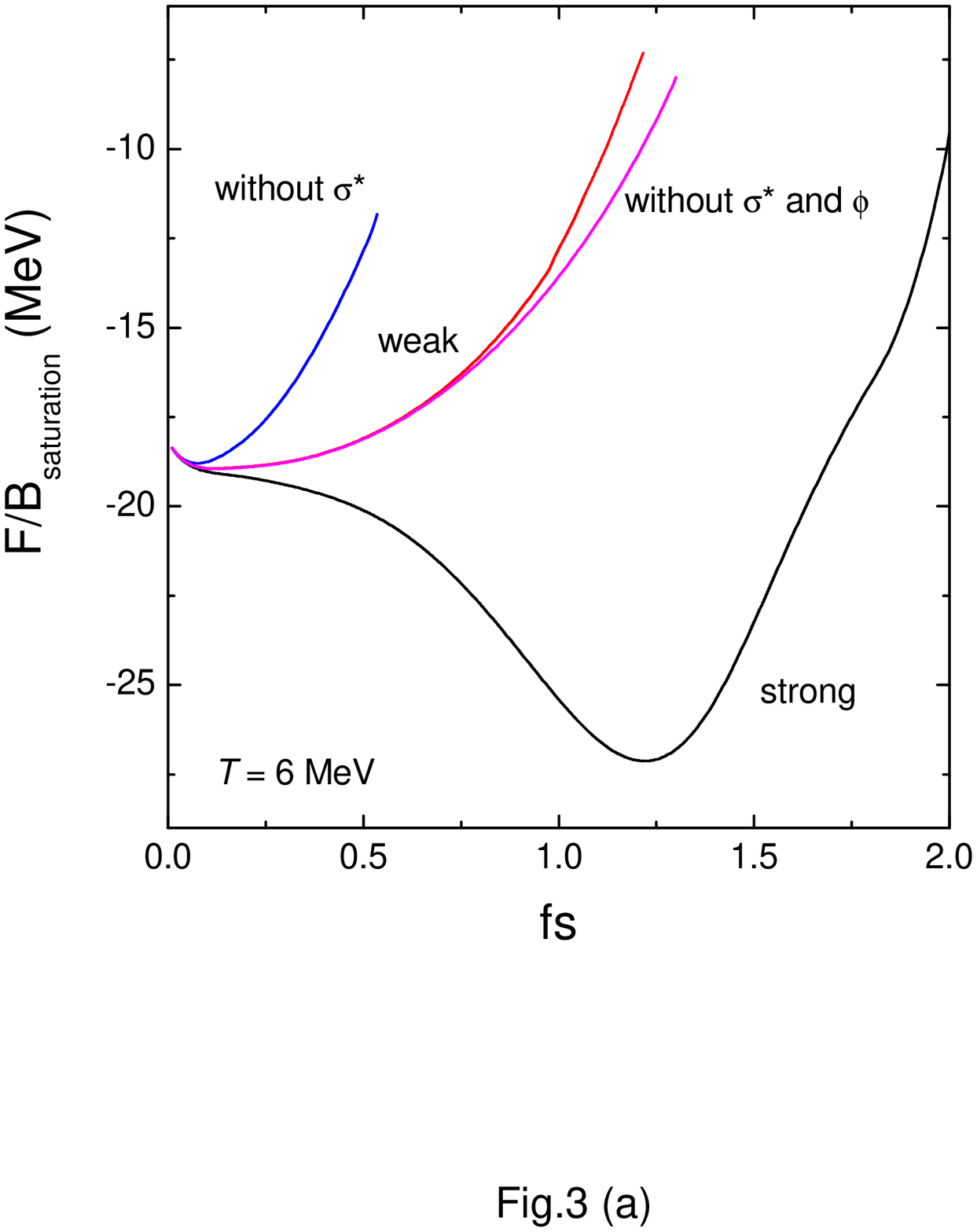}
\includegraphics[width=7cm,height=10cm,angle=0]{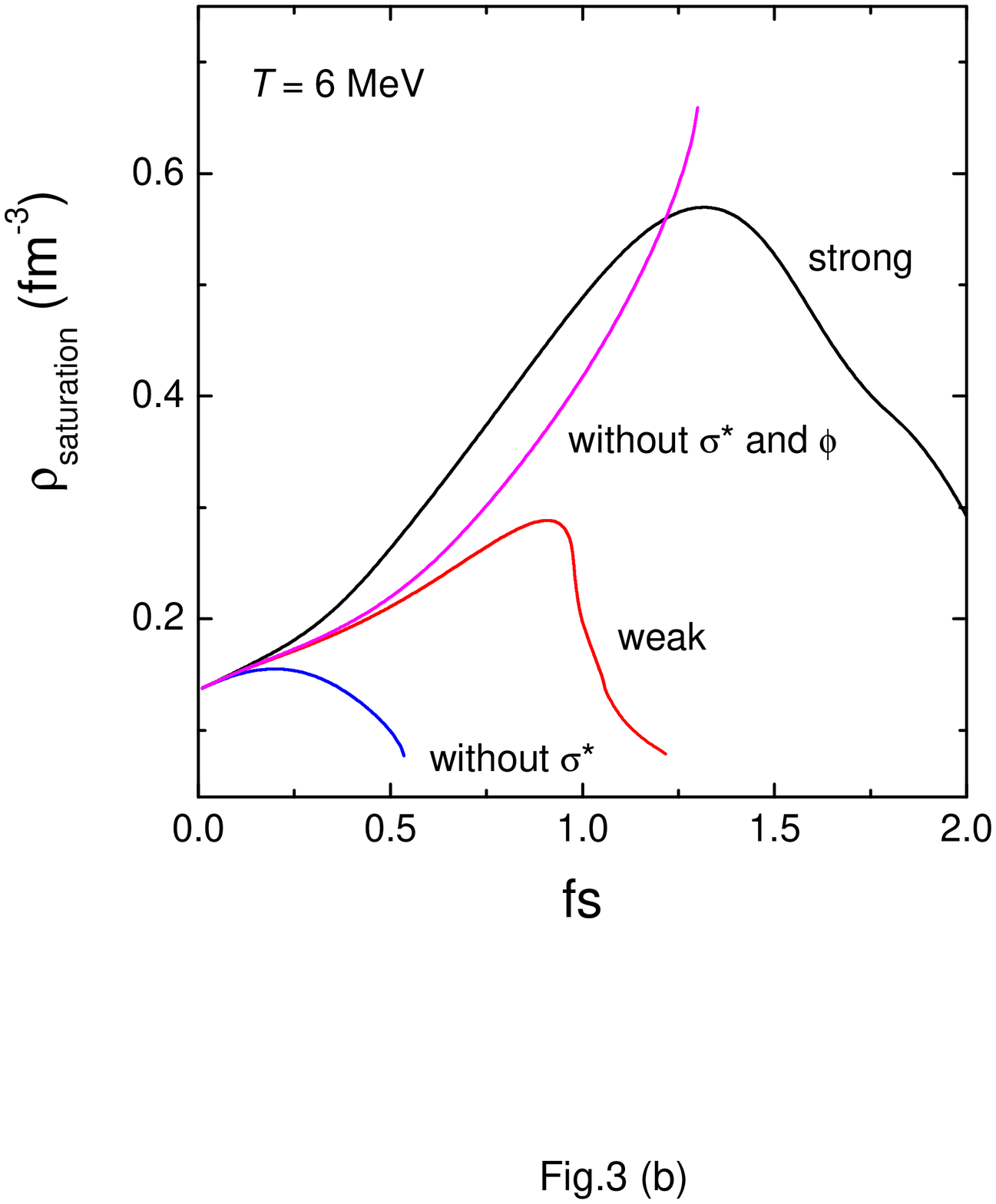}
 \end{center}
\caption{ (a)The minimized free energy per baryon in the strange
hadronic matter;(b)The baryon density corresponding to the
minimized free energy presented in (a), as a function of
strangeness fraction $f_S$.\label{fig3} }
\end{figure}

To see the stability of the system against $f_S$, we minimize the
$F/B$ with respect to $\rho_B$ at each strangeness fraction $f_S$
for both of strong and weak Y-Y interactions, leaving out the
unstable points near zero density. As a function of the
strangeness fraction $f_S$, we present in
Fig.3(a) the minimized $F/B$ , in Fig.3(b) the corresponding baryon density $
\rho_B$. In order to examine the role of the strange mesons, we
have also presented the results without $\sigma^*$ or without both
of $\sigma^*$ and $\phi$ mesons. One can see that for the strong
Y-Y interactions, the minimum free energy which is much deeper
than the that of the ordinary nuclear matter($f_S=0$) appears
around $f_S=1.3$, where the system has almost the highest
density. It means that compared with the ordinal nuclear matter
the system becomes more stable. On the contrary, for the weak Y-Y
interactions, the minimum free energy which is only a little
deeper than the ordinary nuclear matter case appears around
$f_S=0.1$. After this point, the free energy increases
monotonously as baryon density increases and becomes larger than
the value for the ordinary nuclear matter when $f_S>0.45$. It
means that the strange hadronic matter with weak Y-Y interactions
has at most the comparable stability to the ordinary  nuclear
matter in small strangeness fraction $f_S$ region and becomes less
stable than the ordinary nuclear matter in large $f_S$ region. If
the $\sigma^*$ and $\phi$ meson fields are switched off, then the
curve coincides with the weak case in small $f_S$ region and
becomes deeper than the curve for weak case in large $f_S$ region.
The curve without the $\sigma^*$ only grows very quickly with the
strangeness fraction. It means that the $\sigma^*$($\phi$) meson
gives rise the attractive(repulsive) interaction between hyperons.
The above situation is quite similar to that in the MQMC
model\cite{Song}. It means that the above consequence in not model
dependent.
\begin{figure}
\begin{center}
\includegraphics[width=7cm,height=10cm,angle=0]{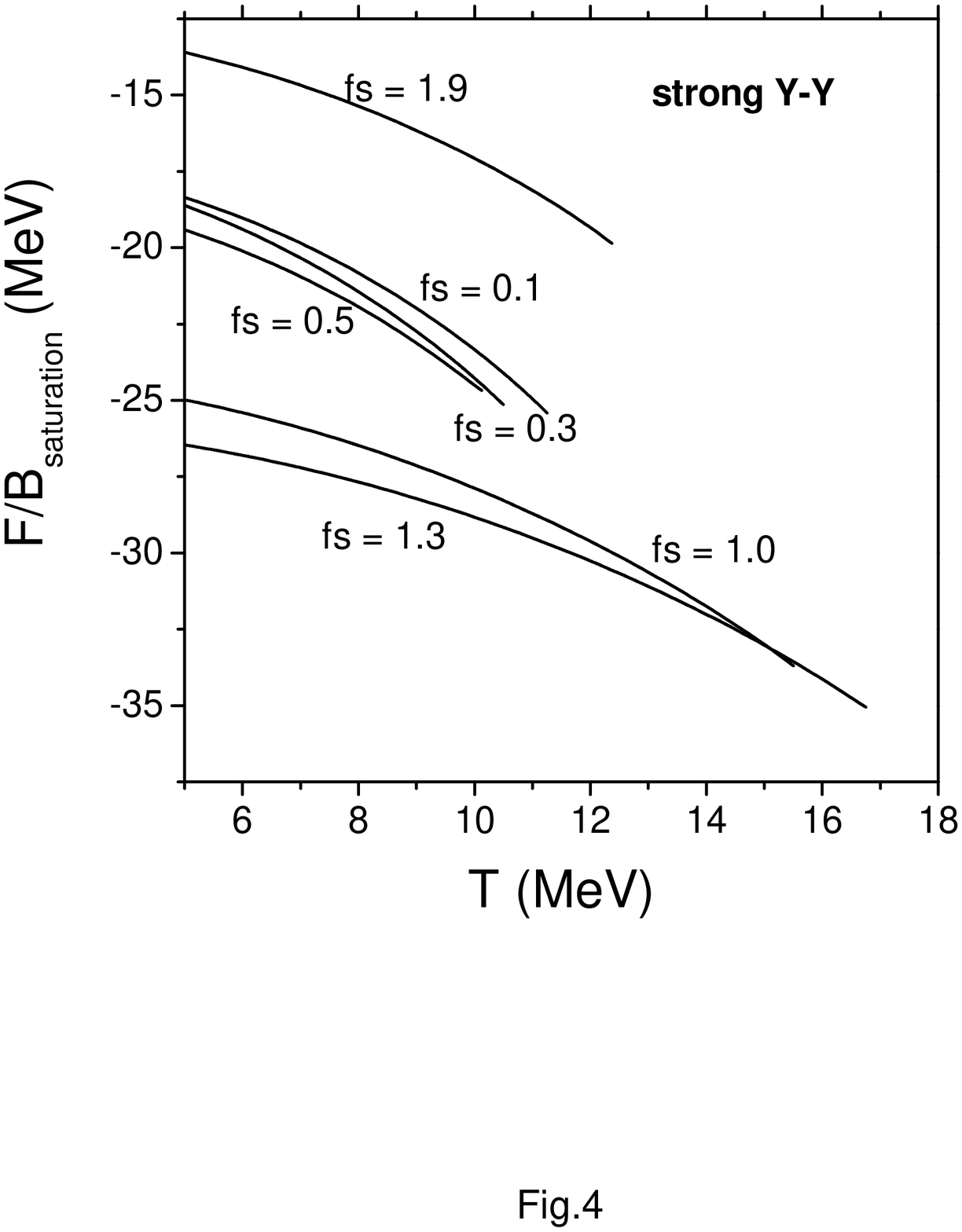}
 \end{center}
\caption{Free energy per baryon at saturation point for the matter
with different $f_S$ as a function of temperature, calculated with
a strong Y-Y interaction.\label{fig4} }
\end{figure}

\begin{figure}
\begin{center}
\includegraphics[width=7cm,height=10cm,angle=0]{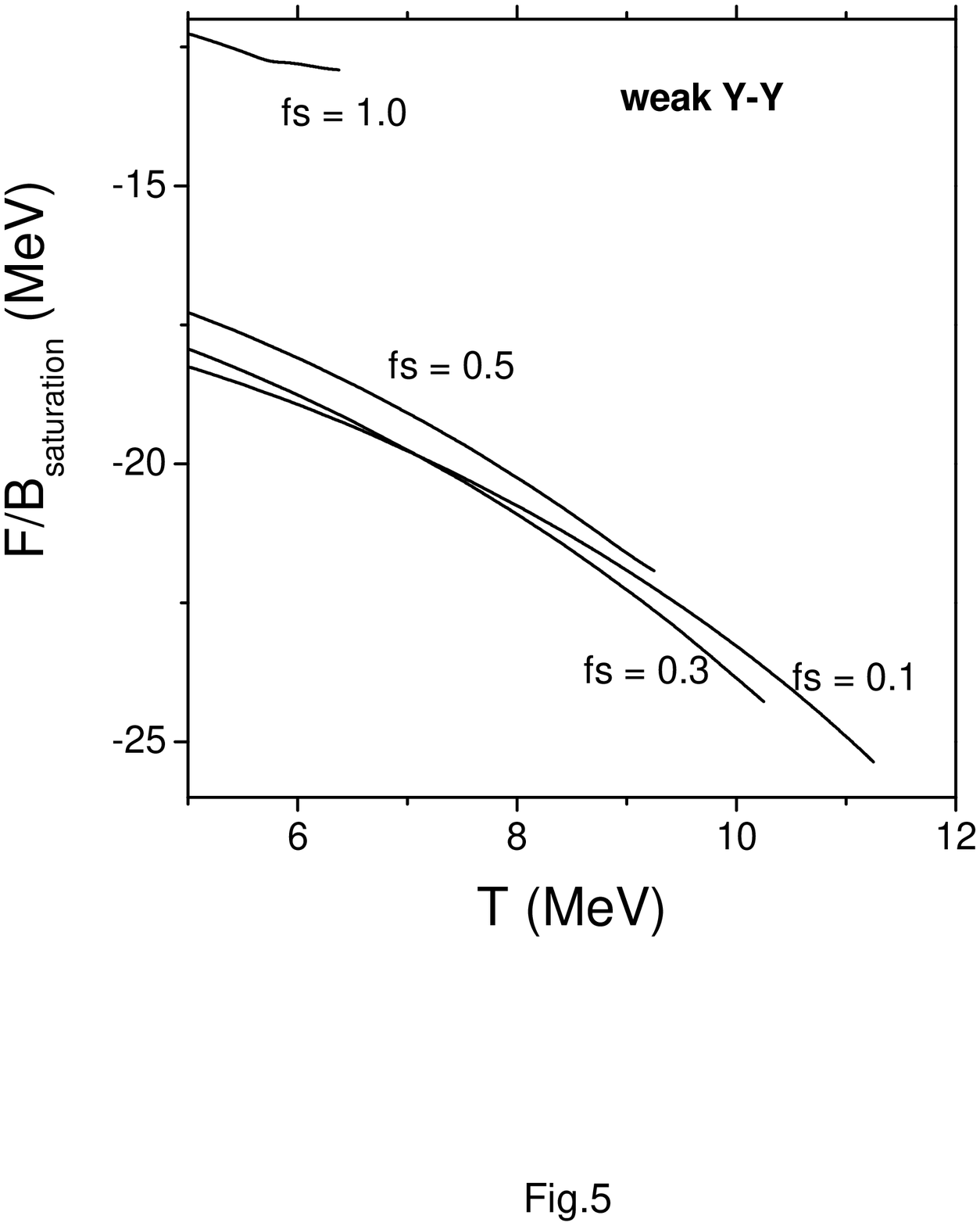}
\end{center}
\caption{The same curves as in Fig.4 but calculated with a weak
Y-Y interaction.\label{fig5} }
\end{figure}

To see the situation of the strange hadronic matter at different
temperature, we present in Fig.4 the free energy at saturation
point verses temperature for the matter with the strong Y-Y
interaction at different strangeness fractions. One can learn that
the matter with different strangeness fraction has different limit
temperature $T_l$, above which the system become unstable, because
the $F/B - \rho$ curves become monotonous and have no minimum when
$T>T_l$. The highest limit temperature appears around $f_S=1.3$
where the system is most stable.  Fig.5 shows the same curves as
in Fig.4 but with the weak Y-Y interactions.  One can see that the
highest limit temperature appears around $f_S=0.1$ where the
system is most stable in this case.
\begin{figure}
\begin{center}
\includegraphics[width=7cm,height=10cm,angle=0]{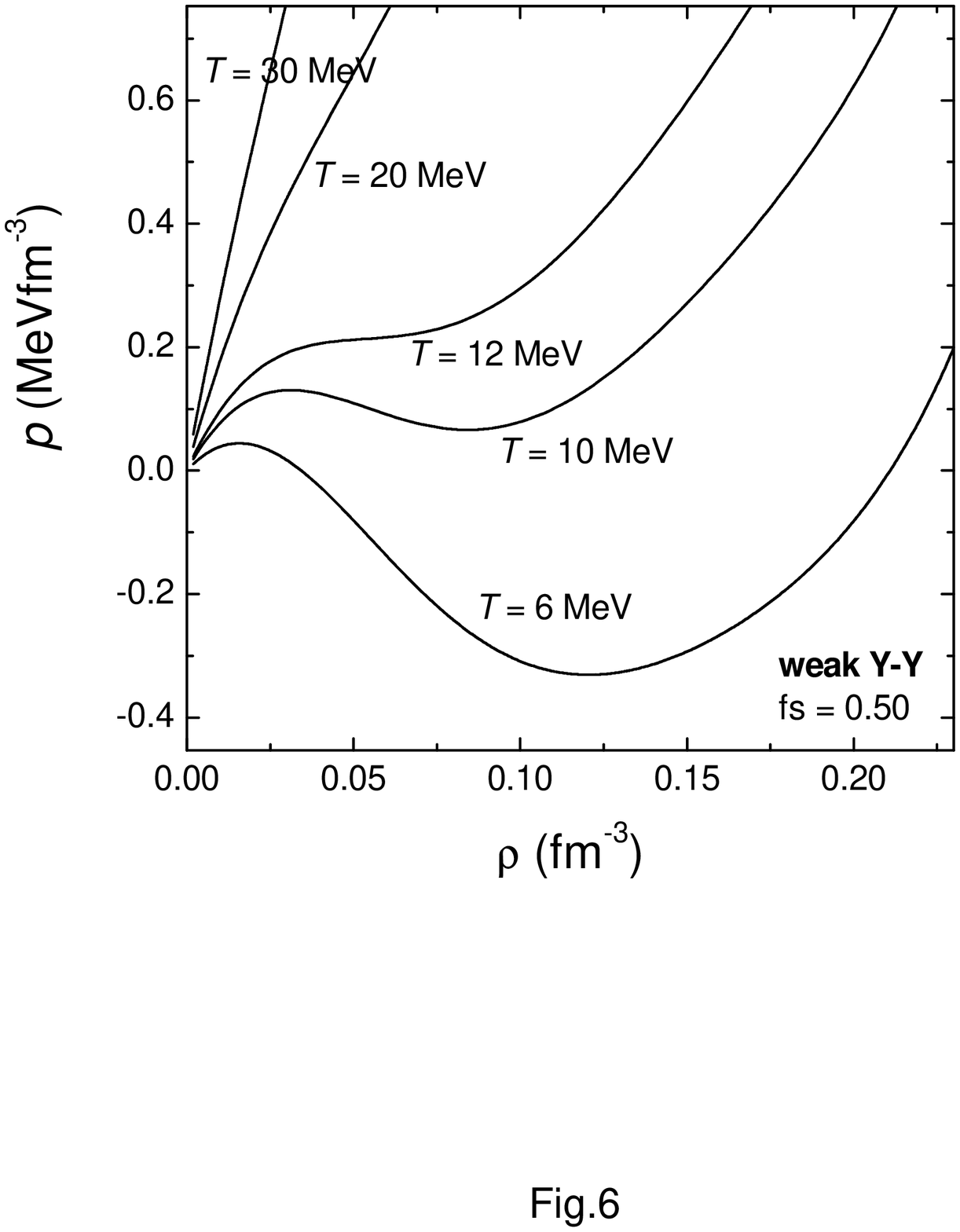}
 \end{center}
\caption{$p-\rho_B$ isotherms of the matter with strangeness
fractions $f_S=0.5$ and at various temperature.\label{fig6} }
\end{figure}
\begin{figure}
\begin{center}
\includegraphics[width=7cm,height=10cm,angle=0]{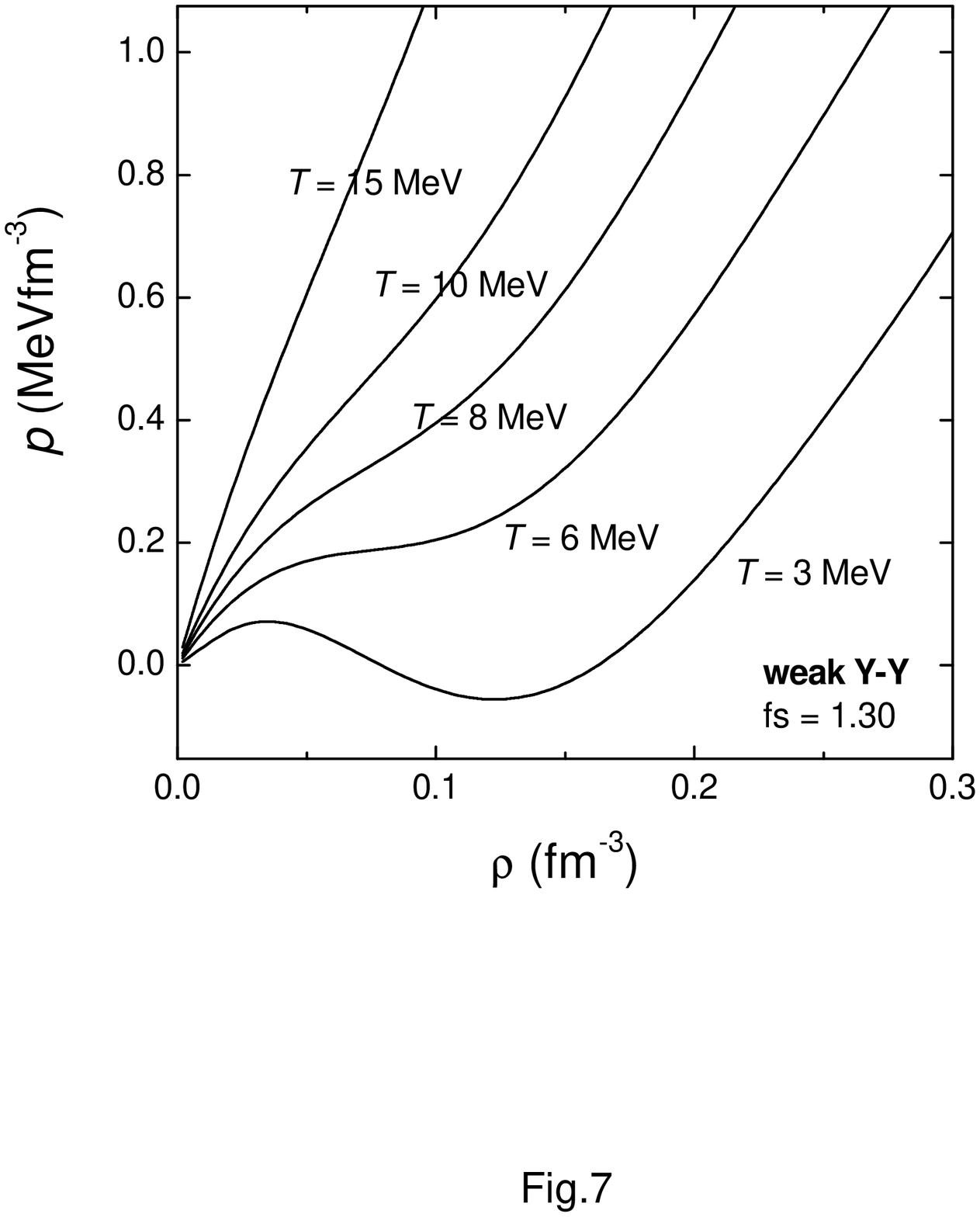}
 \end{center}
\caption{$p-\rho_B$ isotherms of the matter with strangeness fractions $
f_S=1.3$ and at various temperature.\label{fig7} }
\end{figure}

Finally, we discuss the EOS by
plotting the pressure-density ($p-\rho_B)$ isotherms. We show in Fig.6 the $
p-\rho_B$ isotherms of the strange hadronic matter at $f_S=0.5$
and with various temperature. The curve for low temperature
exhibits the same typical shape as given by the Van der Waals
interaction, i.e., there is an unphysical region  where the
pressure decreases with increasing baryon density. An inflection
point appears around $T=12MeV$.
 The situation is
quite similar to the liquid-gas phase transition in ordinary
nuclear matter\cite{Su}. Fig.7 shows the same curves as in Fig.6
but with $f_S=1.3$. In this case, the inflection point appears
around $T=6MeV$. This again indicates the system with large
strangeness fraction  is less stable than the one with small
strangeness fraction when the Y-Y interaction is weak. On
contrary, in the strong Y-Y interaction, the inflection point
appears at a temperature a little higher than 12 MeV for $f_S=0.5$
and around 25 MeV for $f_S=1.3$(see Ref.\cite{Qian}).

In summary, we have extended an effective model describing strange
hadronic matter to finite temperature and then used it to discuss
the properties of multi-hyperon nuclear matter at finite
temperature.  It is found the strange hadronic matter with
different Y-Y interactions behaves very different. While the
system with the strong Y-Y interactions and in a quite large
strangeness fraction region is more stable than the ordinary
nuclear matter, the system with the weak Y-Y interactions is less
stable than the later. This conclusion is not model dependent and
true for both zero and finite temperature.  If the weak
$\Lambda-\Lambda$ interaction is reliable, then the previous
studies on strange hadronic matter and its consequences should be
reexamined.
In particular, if one hope to extend above discussions from infinite strange
hardronic matter to finite hypernuclei, the Coulomb interaction must be considered\cite{Schaffner5}.
It is therefore interesting to perform further
precise measurements of double hypernuclei.

\section*{Acknowledgement}
This work is supported in part by National Natural Science
Foundation of China under No.10375013, 10347107, 10075071, 10247001,
10235030, National Basic Research Program of China 2003CB716300,
the Foundation of Education Ministry of China 20030246005
and CAS Knowledge Innovation Project
N0.KJCX2-N11.  Also supported by the Major State Basic Research
Development Program under contract number G200077400 and
Exploration Project of Knowledge Innovation Program of Chinese
Academy of Sciences.

\end{document}